\documentclass[lettersize,journal]{IEEEtran}
\usepackage{amsmath,amsfonts}
\usepackage[ruled,linesnumbered]{algorithm2e}
\usepackage[caption=false,font=footnotesize,labelfont=rm,textfont=rm]{subfig}
\usepackage{makecell}
\usepackage{stfloats}
\usepackage{ragged2e}
\usepackage{url}
\usepackage{verbatim}
\usepackage{graphicx}
\usepackage{balance}
\usepackage{cite}
\usepackage{multirow}
\usepackage{enumitem}
\usepackage{tikz}
\usetikzlibrary{shapes, arrows.meta, positioning}
\usepackage{booktabs}
\usepackage{lipsum}
\usepackage{amssymb}
\usepackage[colorlinks,
            linkcolor=blue,
            anchorcolor=blue,
            citecolor=blue]{hyperref}
\newcommand{\myrefeq}[1]{(\ref{#1})}
\newcommand{\myreffig}[1]{Fig. \ref{#1}}
\newcommand{\myreftable}[1]{Table \ref{#1}}

\begin{document}
\title{
  IMNet: Interference-Aware Channel Knowledge Map Construction and Localization
  }
\author{
  Le Zhao, Zesong Fei, \textit{Senior Member, IEEE}, Xinyi Wang, \textit{Member, IEEE}, Jingxuan Huang, \textit{Member, IEEE}, \\ Yuan Li, Yan Zhang, \textit{Senior Member, IEEE}
  \thanks{
  Le Zhao, Zesong Fei, Xinyi Wang, Jingxuan Huang, Yuan Li, and Yan Zhang are with the School of Information and Electronics, Beijing Institute of Technology, Beijing 100086, China (e-mail: tobin$\_$bit@icloud.com, feizesong@bit.edu.cn, bit$\_$wangxy@163.com, jxhbit@gmail.com, 3220235334@bit.edu.cn, zhangy@bit.edu.cn).

  Corresponding author: Xinyi Wang
  }
  \vspace*{-0.8cm}
}

\markboth{Journal of \LaTeX\ Class Files,~Vol.~18, No.~9, May~2024}%
{How to Use the IEEEtran \LaTeX \ Templates}

\maketitle
\begin{abstract}
  This paper presents a novel two-stage method for constructing channel knowledge maps (CKMs) specifically for A2G (Aerial-to-Ground) channels in the presence of non-cooperative interfering nodes (INs).
  We first estimate the interfering signal strength (ISS) at sampling locations based on total received signal strength measurements and the desired communication signal strength (DSS) map constructed with environmental topology.
  Next, an ISS map construction network (IMNet) is proposed, where a negative value correction module is included to enable precise reconstruction.
  Subsequently, we further execute signal-to-interference-plus-noise ratio map construction and IN localization. Simulation results demonstrate lower construction error of the proposed IMNet compared to baselines in the presence of interference.
\end{abstract}

\begin{IEEEkeywords}
  Channel knowledge map, deep learning, interfering signal strength, interference localization
\end{IEEEkeywords}

\vspace{-0.2cm}
\section{Introduction}
\IEEEPARstart{W}{ith} the advancement of unmanned aerial vehicle (UAV) technology, traditional ground networks struggle to support the high dynamic characteristics of air-to-ground (A2G) channels. This challenge arises from the rapid mobility of UAVs.
In this context, a novel concept known as the channel knowledge map (CKM) has emerged, garnering significant attention \cite{ZengYong_2021_IEEEWC}.
Among the various channel knowledge, the most widely studied one is channel gain, which can be used to estimate signal-to-noise ratio (SNR) and guide modulation/coding scheme selection \cite{ZengYong_2024_IEEE_ST}, placement optimization \cite{JieXu_2022_WCNC}, as well as 3D-urban environment reconstruction \cite{ChenJunting_2023_TWC}.


Generally, the methods for constructing a CKM can be broadly categorized into two types: model-driven and data-driven methods. Model-driven methods treat channel propagation as a deterministic function containing the effects of blockage and the distance between the transmitter and receiver. Among these, ray tracing stands out as a classic technique \cite{Rizk_1997_TVT}.
While constructing models, the parameters therein can be estimated via minimal mean square error (MMSE), maximum likelihood (ML), least squares (LS), and expectation maximization (EM), etc., based on measurements \cite{Malmirch_2012_TWC, ZengYong_2022_WCNC, XiaoliXu_2024_TWC}.
Although the model-driven methods are able to achieve accurate channel construction, they suffer from high computational complexity and limited models and hence are not suitable for a large-scale network.
Conversely, data-driven methods treat CKM construction as an interpolation problem, employing geostatistical interpolation tools such as kriging \cite{Krige_2017_TVT}, k-nearest neighbors (KNN) \cite{KNN_2009_TIP}, inverse-distance weighting (IDW) \cite{IDW_2012_ISCT}, and radial basis functions (RBF) \cite{Naranjo_2014_WCNC}. Recently, deep neural networks (DNNs) have been leveraged for their robust feature extraction capabilities.
For instance, \cite{Levie_2021_TWC} utilized convolutional neural networks (CNNs) for CKM construction, incorporating environmental topology as a mask matrix input. In \cite{CehnGuokai_2023_CL}, the authors transformed spatially sparse measurements into a graph structure, using graph neural networks to recover the overall CKM. Additionally, \cite{ChenJunting_2024_ICC, ShibliAli_2024_ICC} employed generative adversarial networks (GANs) and diffusion probabilistic models (DPM) to generate unknown CKMs based on sparse measurements and environmental data.

However, existing methods primarily focus on CKM construction for the desired communication link, relying on accurate received signal strength (RSS) measurements at sampling points. In practical scenarios, the presence of interference nodes (INs) can degrade the accuracy of these measurements, leading to reduced fidelity in CKM construction.
To address this issue, we investigate the construction of interfering CKM in the presence of unknown INs in this paper. \textit{To the best of our knowledge, this is the first attempt at constructing interference-aware CKM.} The specific contributions are as follows:

\begin{itemize}[itemsep=2pt,topsep=0pt,parsep=0pt]
  \item We propose a novel two-stage interference-aware CKM construction method. Specifically, interfering signal {strength} (ISS) is firstly estimated with the sampled measurements and the desired communication signal constructed with the environmental information.
  {Subsequently, the ISS map is reconstructed through the proposed ISS map construction network (IMNet), which integrates a negative correction module to encode negative value indicators and enhance map accuracy.}
  \item Utilizing the estimated ISS map, we further construct the signal-to-interference-plus-noise ratio (SINR) map and locate INs. Simulation results show high accuracy in both tasks, demonstrating the effectiveness of our method.
\end{itemize}

The remainder of this paper is organized as follows: Section II introduces the system model and problem formulation. Section III provides details on the IN information extraction and the proposed IMNet. Section IV evaluates the proposed method, and Section V concludes our work.

\vspace{-0.22cm}
\section{System Model}

\begin{figure}[!t]
  \centering
  \includegraphics[width=3.15in]{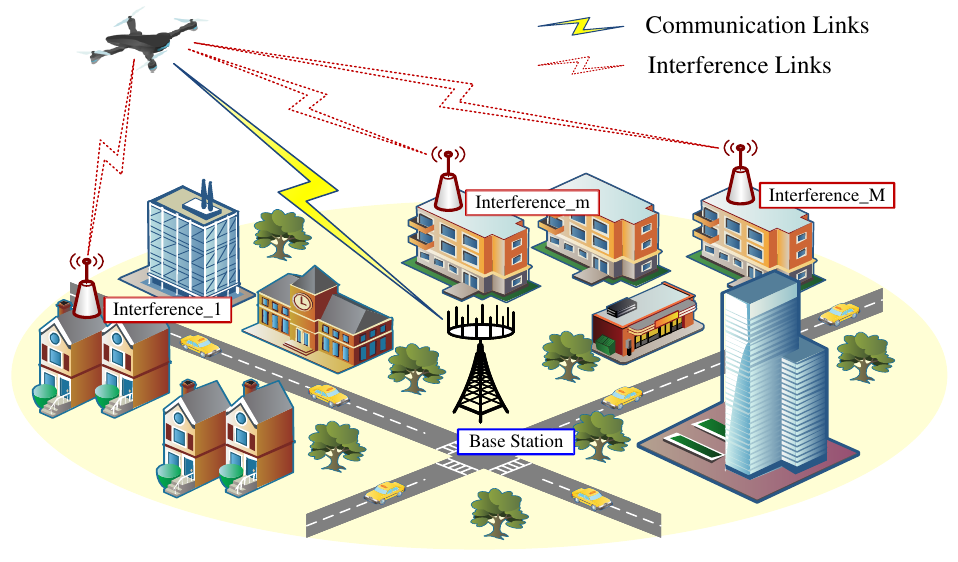}
  \caption{An illustration of cellular-connected UAV communication system with non-cooperative INs.}
  \label{fig.1_System_Model}
  \vspace{-0.5cm}
\end{figure}

We consider the construction of CKM in an urban environment in terms of RSS and SINR, as depicted in \myreffig{fig.1_System_Model}. In this scenario, a cellular-connected UAV maneuvers at a fixed altitude $H_0$ over the area to perform tasks, such as data collection or cargo delivery, while maintaining a connection with a ground base station (GBS).
GBS continuously transmits control signals, while the UAV measures the RSS with received signals and feedback to GBS. GBS then constructs the CKM, which can be used to facilitate UAV trajectory design \cite{JieXu_2022_WCNC}. Additionally, there are $M$ non-cooperative INs operating in the same frequency band, randomly distributed across unknown locations.
{The modeled environment is a three-dimensional rectangular space with dimensions $L$, $W$, and $H$ (length, width, and height, respectively), where $L, W, H \in \mathbb{N}^+$ represent the physical boundaries of the area. Within this space, buildings are randomly distributed, resulting in signal blockages. We denote the set of blockage regions as $\mathcal{D}$.}
Given a resolution of $\Delta \kappa$, with $b_L = L/\Delta \kappa$ and $b_W = W/\Delta \kappa$, we have $\mathbf{Q}_0 \in \mathbb{R}^{b_L \times b_W}$, the location $\mathbf{q}$ at a specific altitude ${H}_0$ is represented as $\mathbf{q} = [x, y]^T \in \mathbf{Q}_0$, where $x$ and $y$ are integers within the ranges $[0, b_L]$ and $[0, b_W]$, respectively.

Assuming the GBS is located in the center of the scenario, with the location denoted as $\mathbf{q}_{\text{BS}}$, the desired communication signal strength (DSS) from GBS to locations $\mathbf{q}$ can be expressed as $R_{\text{BS}}(\mathbf{q}) = p_{\text{BS}} \cdot G(\mathbf{q}, \mathbf{q}_{\text{BS}})$, where $p_{\text{BS}}$ is the GBS's transmit power and $G(\mathbf{q}, \mathbf{q}_{\text{BS}})$ is the channel gain between location $\mathbf{q}$ and $\mathbf{q}_{\text{BS}}$.
The channel gain in [dB] can be expressed with three major components, i.e., the path loss, shadowing, and multi-path fading \cite{XiaoliXu_2024_TWC}, as
 \begin{small}
\begin{align}
  G_{\text{dB}}(\mathbf{q}, \mathbf{q}_{\text{BS}}) = \beta_k + \alpha_k \log_{10} \Vert \mathbf{q}-\mathbf{q}_{\text{BS}} \Vert + \omega(\mathbf{q}) + \upsilon(\mathbf{q}),
  \label{eq.Channel_Define}
\end{align}
\end{small}
where $\alpha_k, \beta_k$ account for the path loss exponent and the path loss intercept, {$\omega(\mathbf{q}) \sim \mathcal{N}(0, \sigma_{\omega}^2)$ models the effect of unpredictable channel shadowing, $\upsilon(\mathbf{q})\sim \mathcal{CN}(0, \sigma_{\upsilon}^2)$ models the effect of multi-path fading, }and $k\in \{0,1\}$ is the indicator of LoS/NLoS condition.
By denoting the location and interfering power of the $m$-th unknown IN as $\mathbf{q}_{\text{IN}}^{m}$ and $p_{\text{IN}}^m$,
{the received interfering signal in location $\mathbf{q}$ can be written as
\begin{small}
	\begin{align}
		Y_{\rm IN}(\mathbf{q})=\sum_{m=1}^{M}{ \sqrt{p^m_{\rm IN} \cdot G(\mathbf{q}, \mathbf{q}_{\rm BS})} X^m_{\rm IN}} + Z,
	\end{align}
\end{small}
where $X^m_{\rm IN}\sim \mathcal{CN}(0, 1)$ is the normalized signal emitted from the $m$th IN,
$Z \sim \mathcal{CN}(0, \sigma^2_z)$ is the Gaussian at the receiver and is thus independent of location $\mathbf{q}$.
Since the INs work in a non-cooperative manner, the average RSS of the interference signals received at location $\mathbf{q}$ can be calculated as
\begin{small}
	\begin{align}
		R_{\text{IN}}(\mathbf{q}) = \mathbb{E} \left[ \Vert Y_{\rm IN}(\mathbf{q})-Z \Vert^2 \right] = \sum_{m=1}^M{p^m_{\rm IN} \cdot G(\mathbf{q}, \mathbf{q}_{\rm BS})}.
	\end{align}
\end{small}
The total RSS received at location $\mathbf{q}$ can be expressed as $R(\mathbf{q}) = R_{\text{BS}}(\mathbf{q}) + R_{\text{IN}}(\mathbf{q})$.
}

According to the above elaboration, the ISS map, containing the RSS of received interference signals at each location,  can be expressed as
\begin{small}
	\begin{align}
		\mathbf{R}_{\text{IN}} = \mathcal{F}_1(\mathbf{\Psi, \Phi}, \mathbf{Q}_{\text{IN}}, \mathbf{P}_{\text{IN}}),
		\label{eq.Interference_model}
	\end{align}
\end{small}
where $\mathbf{\Psi} = \{\mathcal{D}, \mathbf{q}_{\text{BS}}, p_{\text{BS}}\}$, $\mathbf{\Phi} = \{\alpha_k, \beta_k, \sigma_\omega^2, \sigma_\upsilon^2 \}$, $\mathbf{Q}_{\text{IN}} = \{\mathbf{q}^1_{\text{IN}}, \dots, \mathbf{q}^M_{\text{IN}}\}$, $\mathbf{P}_{\text{IN}}=\{p_{\text{IN}}^1,\dots,p_{\text{IN}}^M\}$, and $\mathcal{F}_1(\cdot)$ is the function mapping environment topology and transmitter information to ISS map $\mathbf{R}_{\text{IN}}$.
Intuitively, the total RSS map can be stored as a matrix denoted as $\mathbf{R}\in\mathbb{R}_{+}^{b_L \times b_W}$. Similarly, the DSS map and SINR map can be obtained as $\mathbf{R}_{\text{BS}}\in\mathbb{R}_{+}^{b_L \times b_W}$ and $\mathbf{\Gamma}\in\mathbb{R}_{+}^{b_L \times b_W}$.
The SINR at location $\mathbf{q}$ can be given by
\begin{small}
	\begin{align}
		\Gamma(\mathbf{q}) = \frac{p_{\text{BS}} \cdot G(\mathbf{q}, \mathbf{q}_{\text{BS}})}{R_{\text{IN}}(\mathbf{q}) + \sigma^2_z}.
		\label{eq.SINR_cal}
	\end{align}
\end{small}

The main objective of this paper is to develop a method for constructing ISS map and SINR map as well as localizing INs. To achieve this, we aim to jointly minimize the combined errors in the estimated ISS map, SINR map, and the locations of INs. The problem is mathematically formulated as the following multi-objective optimization problem
	\begin{small}
		\begin{align}
			\underset{\widehat{\mathbf{R}}_{\rm IN},\widehat{\mathbf{\Gamma}},\widehat{\mathbf{Q}}_{\rm IN}}{\min} \left\{\Vert \mathbf{R}_{\text{IN}} - \widehat{\mathbf{R}}_{\text{IN}} \Vert^2_F,  \Vert \mathbf{\Gamma} - \widehat{\mathbf{\Gamma}} \Vert^2_F, \Vert  {\mathbf{Q}}_{\text{IN}} - \widehat{\mathbf{Q}}_{\text{IN}} \Vert^2_F\right\},  \tag{6}
			\label{eq.objective}
		\end{align}
	\end{small}
with the detailed mechanism for estimating $\widehat{\mathbf{R}}_{\text{IN}}$, $\widehat{\mathbf{\Gamma}}$, and $\widehat{\mathbf{Q}}_{\text{IN}}$ to be presented in Section III.



\vspace{-0.2cm}
\section{Interference-aware CKM Construction}
In this section, we first introduce the general workflow of this paper, followed by the data pre-processing and the network architecture of the proposed IMNet. Finally, we estimate the SINR map and the locations of the INs.

\vspace{-0.3cm}
\subsection{General Workflow}

We first denote the limited samples as $\mathbf{R}^S = \mathbf{S} \odot \mathbf{R}$, where $\odot$ denotes the Hadamard matrix product, and $\mathbf{S}$ is the measurement indicator matrix with elements ${S}(\mathbf{q}) \in \{0,1\}$. The sampling location set is denoted as $\mathbf{Q}^S$.
The estimation of parameters $\{\hat{\alpha}, \hat{\beta}\}$ of the channel model in \myrefeq{eq.Channel_Define} has been studied in \cite{XiaoliXu_2024_TWC}.
{Hence, in this paper, we only estimate the path loss-related parameters, denoted as $\hat{\alpha}_k$ and $\hat{\beta}_k$. Nevertheless, while evaluating the performance of the proposed model, both channel shadowing and multi-path fading are taken into account for generating RSS samples.} The estimated channel parameters are represented as $\widehat{\mathbf{\Phi}} = \{\hat{\alpha}_k, \hat{\beta}_k, 0, 0\}$. This allows us to construct the channel gain corresponding to GBS as $\widehat{G}(\mathbf{q},\mathbf{q}_{\text{BS}})$ and the DSS at location $\mathbf{q}$ as $\widehat{R}_{\text{BS}}(\mathbf{q}) = p_{\text{BS}} \cdot \widehat{G}(\mathbf{q}, \mathbf{q}_{\text{BS}}) \in \widehat{\mathbf{R}}_{\rm BS}$, using a well-trained deep learning network \cite{Levie_2021_TWC}.
However, estimating the locations and transmit power of the IN remains challenging, and $\mathcal{F}_{1}$ in \myrefeq{eq.Interference_model} is unavailable anymore. {Therefore, we extract the ISS at the sampling points first, based on the constructed DSS map and the sampled mixture signal strength, i.e.,  $\widehat{R}^S_{\rm IN}(\mathbf{q})=R^S(\mathbf{q})-\widehat{R}_{\rm BS}(\mathbf{q})$. And then construct the ISS map.} The overall procedure can be expressed as
\begin{small}
	\begin{align}
		\widehat{\mathbf{R}}_{\text{IN}} &= \mathcal{F}_{2}(\mathbf{\Psi}, \widehat{\mathbf{\Phi}}, \mathbf{R}^S).
		\label{eq.Interference_func}
	\end{align}
\end{small}
It is important to note that \(\mathcal{F}_2(\cdot)\) differs from \(\mathcal{F}_1(\cdot)\) as it leverages \(\mathbf{R}^S\) to capture IN information, rather than relying on the prior information in \myrefeq{eq.Interference_model}. Additionally, \(\mathcal{F}_2(\cdot)\) represents the ISS map estimator which needs to be learned by the proposed IMNet.
After obtaining $\widehat{\mathbf{R}}_{\rm IN}$, the corresponding SINR at location $\mathbf{q}$ can be estimated as
\begin{small}
	\begin{align}
		\widehat{\Gamma}(\mathbf{q}) = \frac{p_{\text{BS}}\cdot \widehat{G}(\mathbf{q},\mathbf{q}_{\text{BS}})}{\widehat{R}_{\text{IN}}(\mathbf{q}) + \sigma^2_z},
		\label{eq.SINR_estimation}
	\end{align}
\end{small}
with $\hat{R}_{\rm IN}(\mathbf{q})$ being the estimated ISS at location $\mathbf{q}$.
After that, we are able to localize the IN, denoted as $\widehat{\mathbf{Q}}_{\text{IN}}$ by detecting the peaks in $\widehat{\mathbf{R}}_{\text{IN}}$.

\begin{figure*}[!t]
  \centering
  \includegraphics[width=6.5in]{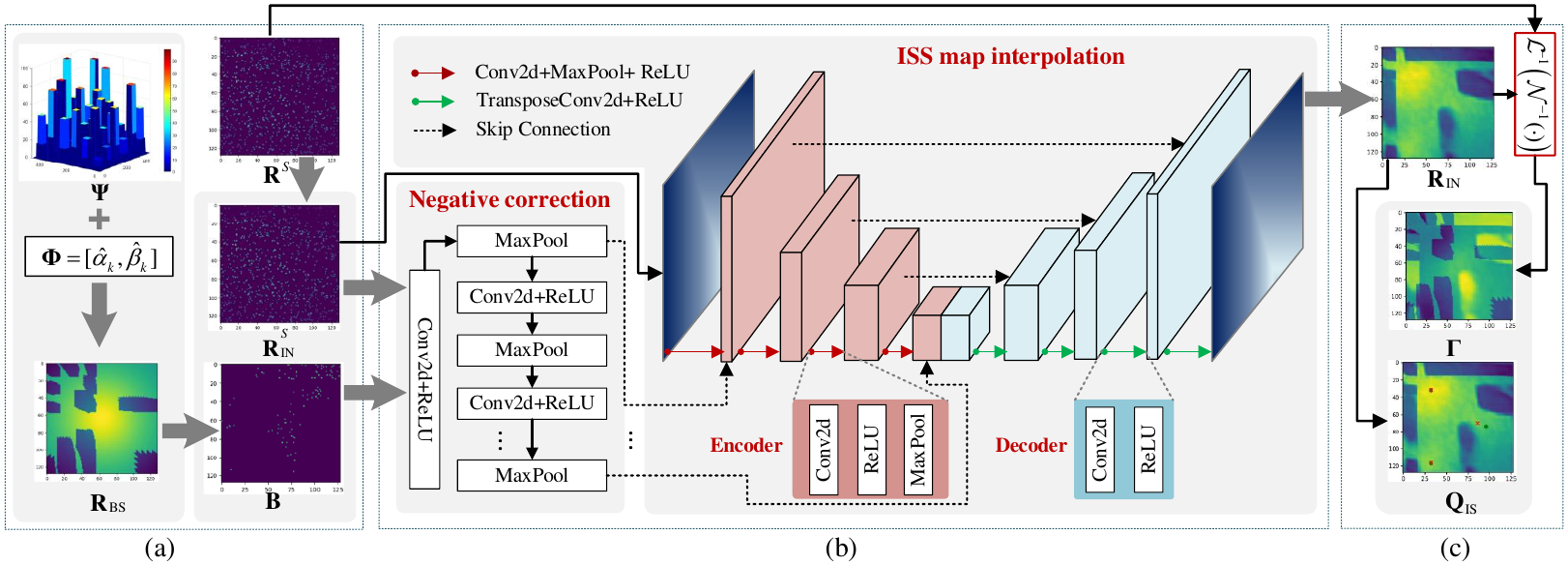}
  \caption{General Workflow}
  \label{fig.2_overall_architecture}
  \vspace{-0.6cm}
\end{figure*}

\vspace{-0.3cm}
\subsection{Pre-processing}

In this paper, we propose to use a DNN model to approximate the function $\mathcal{F}_2$, as an estimator for an ISS map construction.
Without loss of generality, each RSS measurement is firstly transformed to an image using logarithm operation and normalization, while the unsampled areas are initialized as $0$.
Here we use the $\mathcal{L}(\cdot)$ and $\mathcal{N}(\cdot)$ to denote the logarithm operation and normalization for simplicity.
To construct the complete ISS map, we need to extract the interfering RSS from the sampled ${\mathbf{R}}^S$ first.
With the estimated channel parameters $\widehat{\mathbf{\Phi}}$ and the environment $\widehat{\mathbf{\Psi}}$, we obtain the sampled ISS map estimaion, denoted as $\widehat{\mathbf{R}}_{\text{IN}}^S = {\mathbf{R}}^S - \mathbf{S} \odot \widehat{\mathbf{R}}_{\text{BS}}$.
However, due to the existence of noise and measurement error, there may exist many \textit{negative} RSS values in $\widehat{\mathbf{R}}_{\text{IN}}^S$, resulting in infinite $\mathcal{L}(\widehat{\mathbf{R}}_{\text{IN}}^S)$ and severe degradation in CKM construction.
{Therefore, it is necessary to preprocess $\widehat{\mathbf{R}}_{\text{IN}}^S$, corresponding to \myreffig{fig.2_overall_architecture}(a).}
We first define the negative indicator matrix to indicate the negative position to help the subsequent step to complete the negative power correction, denoted as $\mathbf{B} = \mathbb{I}( \widehat{\mathbf{R}}^S_{\rm IN} < 0 )\in\mathbb{R}_{+}^{b_L \times b_W}$.
{Next, we take the absolute values of the negative entries in $\widehat{\mathbf{R}}^S_{\text{IN}}$ by converting them to their opposites, denoted as $\widehat{\mathbf{R}}^{S,+}_{\text{IN}}$, enabling a finite logarithmic representation.
The pre-processed sampling measurements is denoted as $\overline{\mathbf{R}}^S_{\rm IN}=\mathcal{N}(\mathcal{L}(\widehat{\mathbf{R}}^{S,+}_{\text{IN}}))$.}

\vspace{-0.3cm}
\subsection{IMNet Architecture}

The proposed IMNet, depicted in \myreffig{fig.2_overall_architecture}(b), comprises two key modules: {negative correction module} and ISS map interpolation module.
Specifically, in the negative correction module, {both pre-processed sparse ISS map measurements $\overline{\mathbf{R}}^S_{\rm IN}$ and the corresponding negative indicator matrix $\mathbf{B}$,} are passed through a series of convolutional layers for feature extraction and negative correction. The output is then skip-connected with the feature maps of the ISS map interpolation module, where the UNet architecture is utilized for image interpolation \cite{Levie_2021_TWC}.
The UNet structure consists of an encoder path with convolutional layers and ReLU activation to generate feature maps, followed by a decoder path that up-samples these feature maps.
In particular, skip connections between corresponding encoder and decoder layers preserve high-resolution features, contributing to accurate reconstruction. The output is represented as $\overline{\mathbf{R}}_{\text{IN}}$, a normalized matrix in the logarithm domain.
{With an input feature size $b_L \times b_W$, the floating-point operations (FLOPs) of each 2D convolutional layer is $\mathcal{O}(K^2C_{\rm in}C_{\rm out}b_Lb_W)$, where $K$ is the number of pixels of each filter kernel along the $x$, $y$ axis, $C_{\rm in}$ and $C_{\rm out}$ are the input and output channels. }

\vspace{0.08cm}
IMNet is trained using a loss function based on the Mean Squared Error (MSE) between the processed ISS map $\mathcal{N}(\mathcal{L}(\mathbf{R}_{\text{IN}}))$ and $\overline{\mathbf{R}}_{\text{IN}}$, defined as
\begin{small}
	\begin{align}
		\text{MSE} = \frac{1}{\vert \mathbf{Q}_0 \vert} \left\Vert \mathcal{N}(\mathcal{L}(\mathbf{R}_{\text{IN}}))-\overline{\mathbf{R}}_{\text{IN}} \right\Vert^2_F.
		\label{eq.MSE_define}
 	\end{align}
\end{small}
This completes the ISS map reconstruction process.

\vspace{-0.2cm}
\subsection{SINR Construction and IN Localization}

As depicted in \myreffig{fig.2_overall_architecture}(c), we execute inverse normalization and transform the IMNet output $\overline{\mathbf{R}}_{\text{IN}}$ back to a linear scale to obtain the SINR map $\widehat{\mathbf{\Gamma}}$.
Specifically, we extract the non-negative RSS as \(\widehat{\mathbf{R}}^+_{\text{IN}} = \widehat{\mathbf{R}}_{\text{IN}}^S - \mathbf{B} \odot \widehat{\mathbf{R}}^S_{\text{IN}}\), and apply \(\mathcal{L}(\cdot)\) to transform \(\widehat{\mathbf{R}}^+_{\text{IN}}\) into the logarithmic domain. The vectorized form of \(\mathcal{L}(\widehat{\mathbf{R}}^+_{\text{IN}})\) is denoted as \(\hat{\mathbf{r}}^+ = [\hat{r}^+_1,\dots,\hat{r}^+_{N}]^T\), where \(N = \Vert \mathbf{S-B} \Vert_1\).
We then extract the corresponding RSS values from \(\overline{\mathbf{R}}_{\text{IN}}\) as \(\overline{\mathbf{R}}^+_{\text{IN}} = (\mathbf{S}-\mathbf{B}) \odot \overline{\mathbf{R}}_{\text{IN}}\), and further vectorize \(\overline{\mathbf{R}}^+_{\text{IN}}\) as \(\overline{\mathbf{r}}^+ = [\overline{r}^+_1,\dots,\overline{r}^+_N]^T\). Through fine estimation, we ensure that the MSE in \myrefeq{eq.MSE_define} is close to zero. Denoting the upper and lower bounds of the logarithmic ISS map values as \([r_{\max}, r_{\min}]\), we approximate
\begin{align}
\hat{r}^+_n = \mathcal{N}^{-1}(\overline{r}^+_n) = (r_{\max}-r_{\min})\overline{r}^+_n + r_{\min},
\label{eq.normalization}
\end{align}
where \( n\in \mathbb{Z}, 1 \leq n \leq N\).
We define the least squares (LS) estimation of the coefficients in \myrefeq{eq.normalization} as:
\begin{align}
\arg \underset{\{r_{\max}, r_{\min}\}}{\min} \Vert \hat{\mathbf{r}}^+ - (r_{\max}-r_{\min}) \overline{\mathbf{r}}^+ - r_{\min}\mathbf{1}_{N \times 1} \Vert^2.
\end{align}
Defining \(\mathbf{A} = [\overline{\mathbf{r}}^+ \,\,\, \mathbf{1}_{N \times 1}]\), the LS estimation is then given by:
\begin{align}
\begin{bmatrix}
\hat{r}_{\max}-\hat{r}_{\min} \\ \hat{r}_{\min}
\end{bmatrix}
= (\mathbf{A}^T\mathbf{A})^{-1}\mathbf{A}^T \hat{\mathbf{r}}^+.
\end{align}
The estimated ISS at location $\mathbf{q}$ is subsequently obtained as $\widehat{R}^+_{\rm IN}(\mathbf{q}) = \mathcal{L}^{-1}\left( \bar{R}^+_{\rm IN}(\mathbf{q})(r_{\max} - r_{\min}) + r_{\min} \right)$.
Finally, the SINR map \(\widehat{\mathbf{\Gamma}}\) can be derived based on \myrefeq{eq.SINR_estimation}.

Since the RSS power of the GBS has been removed in $\widehat{\mathbf{R}}_{\text{IN}}$, the RSS will be highest at locations closest to the IN within a given altitude plane. Therefore, we can apply a two-dimensional constant false alarm rate (2D-CFAR) detection algorithm \cite{LiuCong_2021_ICMA} to locate the 2D locations of the INs, denoted as $\widehat{\mathbf{Q}}_{\text{IN}}=\{\hat{\mathbf{q}}_{\text{IN}}^1, \dots, \hat{\mathbf{q}}_{\text{IN}}^{\hat{M}},\}$, where $\hat{M}$ is the number of estimated IN location.

\section{Numerical Results}
\subsection{Simulation Settings}

\begin{figure*}[!t]
  \centering
  \subfloat[ISS maps]{\includegraphics[width=6.2in]{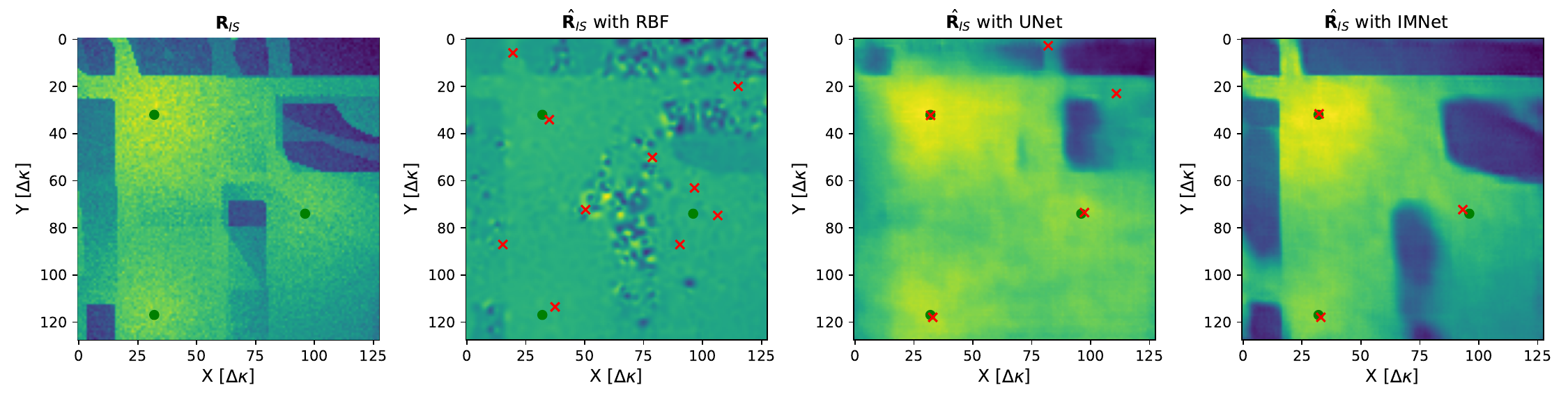}}
  \vspace{-0.2cm}
  \subfloat[SINR maps]{\includegraphics[width=6.14in]{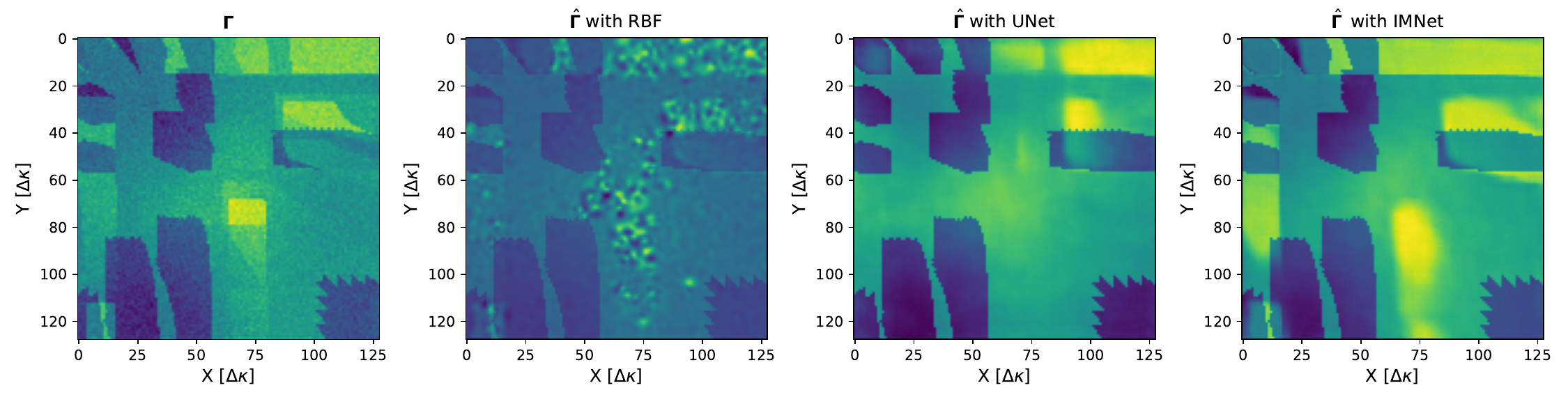}}
  \caption{Reconstructed ISS maps with estimated IN location and SINR maps with different methods.}
\label{fig.3_ReconstructedMaps}
\vspace{-0.5cm}
\end{figure*}

\begin{table}[!t]
  \renewcommand{\arraystretch}{0.9}
  \setlength{\tabcolsep}{0.1mm}
  \centering
  \caption{Structure of the Negative Correction Module}
  \begin{tabular}{c|c|c|c|c|c|c}
    \Xhline{1.1px}
    \,\qquad \textit{Layer\qquad}   & \,\,\,{Input}\,\,\,  & \,\,\,\quad {1\quad}\,\,\,  &\,\,\, \quad {2\quad}\,\,\,  & \,\,\,\quad {3\quad} \,\,\, &\,\,\, \quad {4\quad} \,\,\, & \,\,\, \quad{5\quad}\,\,\, \\
    \hline
    \textit{Resolution} & {128} & {64} & {32} & {16} & {8} & {4} \\
    \textit{Channels} & {1} & {20} & {30} & {40} & {90} & {150}  \\
    \textit{Filter} & {3} & {3} & {3} & {3} & {3} & {3}  \\
    \Xhline{1.1px}
  \end{tabular}
  \label{Table.CNN_structure}
  \vspace{-0.1cm}
\end{table}

\begin{table}[!t]
\renewcommand{\arraystretch}{0.9}
\setlength{\tabcolsep}{0.1mm}
\centering
\caption{Structure of the ISS map Interpolation Module}
\begin{tabular}{c|c|c|c|c|c|c|c|c|c}
  \Xhline{1.1px}
  \qquad \textit{Layer\qquad}  & \,{Input}\,  & \quad {1\quad}  & \quad {2\quad}  & \quad {3\quad}  & \quad {4\quad} & \quad{5\quad} & \quad{6\quad} & \quad{7\quad}  & \quad{8\quad}\quad\\
  \hline
  \textit{Resolution} & {128} & {128} & {64} & {32} & {32} & {16} & {16} & {8} & {4} \\
  \textit{Channels} & {2} & {6} & {40} & {50} & {60} & {100} & {120} & {150} & {320}\\
  \textit{Filter} & {3} & {5} & {5} & {5} & {5} & {3} & {5} & {5} & {4}\\
  \hline

  \textit{Layer} & {9} & {10} & {11} & {12} & {13} & {14} & {15} & {16} & {17}\\
  \hline
  \textit{Resolution} & {4} & {8} & {16} & {16} & {32} & {32} & {64} & {128} & {128}\\
  \textit{Channels} & {300} & {240} & {200} & {120} & {100} & {81} & {27} & {21} & {1}\\
  \textit{Filter} & {4} & {5} & {5} & {3} & {5} & {5} & {5} & {5} & {3}\\
  \Xhline{1.1px}
\end{tabular}
\label{Table.UNet_structure}
\vspace{-0.2cm}
\end{table}

The scenario used for the evaluation of construction methods consists of a network of 1 GBS and $M=3$ INs.
Referring to the International Telecommunication Union (ITU) \cite{ITU_building_distribution_2012}, buildings are generated with three parameters: 1) $a$: the ratio of the area covered by buildings to the whole land area; 2) $b$: the average number of buildings per square kilometer; 3) $\lambda$: the mean value of the building height on Rayleigh distribution.
{The overall square area is $L=W=512$ meters, and the UAV maneuvers at the altitude $H=120$ meters. }
Under the resolution of $\Delta \kappa = 4$ meters, we have $b_W=b_L=128$.
In this paper, we have $a=0.25$, $b=144$, $\lambda=40$.
The channel model parameters in \myrefeq{eq.Channel_Define} are set as $\alpha_0,\beta_0 = (-22,-28)$, $\alpha_1$, $\beta_ 1=(-28,-24)$, $\sigma^2_\omega + \sigma^2_\upsilon=2$, referring to \cite{XiaoliXu_2024_TWC, ChenJunting_2023_TWC}. {The variance of receiver noise is $\sigma^2_z = 10^{-14}$}.
{The GBS transmit power is set as $p_{\text{BS}}=40$ watt, and the interfering power of INs are set as $\mathbf{P}_\text{IN} = \{40, 10, 10\}$ watt.}


To train the model, we generated 1,000 urban maps, along with the corresponding original total RSS map $\mathbf{R}$, DSS map $\mathbf{R}_{\text{BS}}$, ISS map $\mathbf{R}_{\text{IN}}$, and locations of INs $\mathbf{Q}_{\text{IN}}$.
The dataset was split into 700 maps for training, 100 for validation, and 200 for testing\footnote{{The dataset will be available at [https://github.com/tobinzhao/Interfering-CKM-construction] after publication.}}.
During training, the sampling matrix $\mathbf{S}$ was generated randomly. After preprocessing, the MSE between the model output and $\mathbf{R}$ was calculated using \myrefeq{eq.MSE_define}, and model parameters were refined through backpropagation. The number of training epochs is set as 100.

\vspace{-0.3cm}
\subsection{Performance Evaluation}

In this subsection, we evaluate the performance of our proposed algorithm on our dataset.
We employ a 6-layer CNN for negative correction and a 17-layer U-Net for ISS map interpolation.
The detailed structures are provided in \myreftable{Table.CNN_structure} and \myreftable{Table.UNet_structure}.
{The baselines consist of the UNet \cite{Levie_2021_TWC}, which utilizes a CNN structure as detailed in \myreftable{Table.UNet_structure}, alongside traditional spatial interpolation methods, including Kriging \cite{Krige_2017_TVT}, KNN \cite{KNN_2009_TIP}, IDW \cite{IDW_2012_ISCT}, and RBF \cite{Naranjo_2014_WCNC}. For these baselines, the estimated ISS map measurements $\widehat{\mathbf{R}}^S_{\rm IN}$ serve as the input. In contrast, the proposed IMNet takes both $\widehat{\mathbf{R}}^S_{\rm IN}$ and the negative indicator matrix $\mathbf{B}$ as inputs to reconstruct the ISS map.}

\myreffig{fig.3_ReconstructedMaps} shows the construction results of RBF, U-Net, and proposed IMNet. In the ISS maps, the green dots are used to mark true IN locations, and the red crosses are used to mark the estimated results. All the plots were obtained at the sampling rate of 20\%. We can intuitively see that both RBF and UNet suffer from false alarms and high localization errors in large-scale ISS map construction, which further leads to errors in the localization of INs and SINR maps.

To comprehensively evaluate the performance of our proposed method, we consider two key metrics under different sampling rates: normalized mean squared error (NMSE) and localization distance error. Both metrics are averaged over all $N_{\text{test}}=200$ test maps.
In particular, NMSE is used to quantify the accuracy of the reconstructed ISS map and SINR maps are defined as the mean squared error between the estimation $\widehat{\mathbf{R}}_{\rm IN}$, $\widehat{\mathbf{\Gamma}}$ in [dB] with ground truth $\mathbf{R}_{\rm IN}$, and $\mathbf{\Gamma}$ in [dB], respectively. The localization error measures the accuracy of the IN localization while accounting for the possibility of false alarms during peak detection. Specifically, the localization error is defined as
\vspace{-0.1cm}
\begin{small}
	\begin{align}
		\text{Error} = \frac{1}{N_{\text{test}}}\sum_{m=1}^{N_{\text{test}}} \frac{ \sum_{i=1}^{\hat{M}} \underset{n\in\{1,\dots,M\}}{\min} \Vert \hat{\mathbf{q}}^i_{\text{IN}} - \mathbf{q}^m_{\text{IN}} \Vert }{\hat{M}}.
		\label{eq.error_define}
	\end{align}
\end{small}
\vspace{-0.1cm}

\begin{figure*}[!t]
  \centering
  \subfloat[]{\includegraphics[width=1.7in]{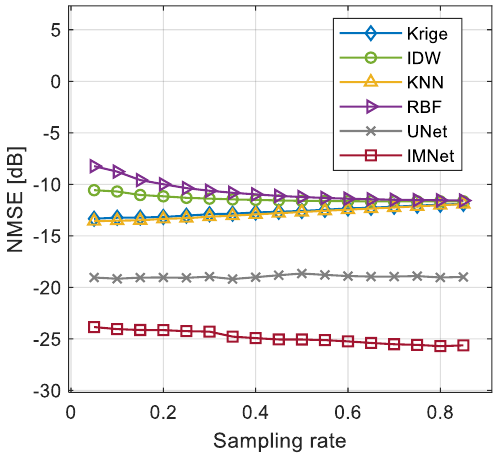}}
  \hfil
  \subfloat[]{\includegraphics[width=1.7in]{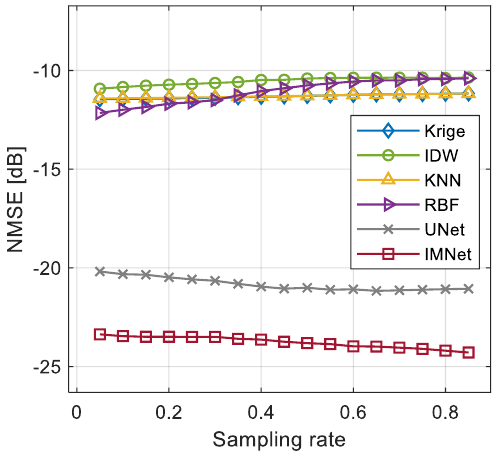}}
  \hfil
  \subfloat[]{\includegraphics[width=1.7in]{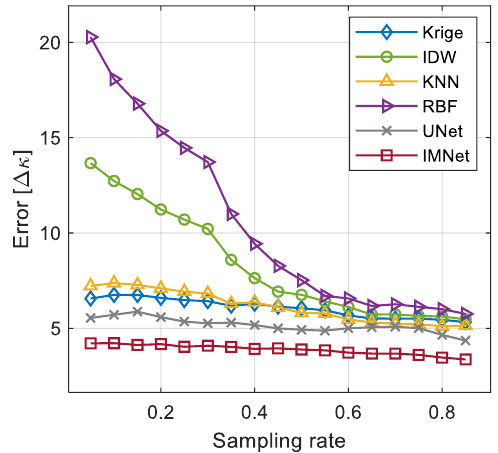}}
  \hfil
  \caption{Construction / Estimation Error for (a) ISS map, (b) SINR map, and (c) IN localization.}
\label{fig.metric_result}
\vspace{-0.6cm}
\end{figure*}

\myreffig{fig.metric_result} illustrates the construction errors across different methods.
In \myreffig{fig.metric_result}(a), the NMSE of ISS map reconstruction is presented. It can be observed that the proposed IMNet consistently achieves lower NMSE compared to other methods across all sampling rates. Moreover, increasing the sampling rate further reduces the reconstruction error. The other methods, however, struggle to handle negative values, resulting in a plateau in reconstruction accuracy even as the measurement rate increases.

\myreffig{fig.metric_result}(b) displays the NMSE of SINR map construction. The results demonstrate that IMNet offers superior construction accuracy compared to the others.
However, due to channel parameter estimation and ISS map construction errors, the SINR estimation error for IMNet increases by approximately 2dB compared to the ISS map reconstruction error.
{Note that, interpolation methods like RBF do not exhibit decreased error with higher sampling rates. This is because these methods are increasingly affected by negative ISS estimates as the sampling rate increases, particularly in high-SINR areas. This leads to higher errors at higher sampling rates.}

Finally, in \myreffig{fig.metric_result}(c), we depict the localization errors calculated using \myrefeq{eq.error_define}. The results show that IMNet consistently provides better localization accuracy over a wide range of sampling rates. The average localization error distance is 4--5 $\Delta \kappa$, indicating that in a 512$\rm{m} \times$512$\rm{m}$ environment, our error is less than 20m.

\vspace{-0.3cm}
\section{Conclusion}
\vspace{-0.1cm}
In this paper, we introduce a novel two-stage method for constructing interference-aware CKM in the presence of unknown non-cooperative INs. By utilizing the total RSS at sampled locations and estimating the DSS map based on environmental topology, we extract the ISS, construct the ISS map, and subsequently estimate SINR maps while localizing the INs. Simulation results show that our method significantly outperforms baseline approaches in these tasks. {For future work, we will investigate SINR lower bound maps using the statistical distributions of shadowing and multi-path fading to better illustrate worst-case coverage.}

\vspace{-0.2cm}
\bibliographystyle{IEEEtran}
\bibliography{IEEEabrv,ZhaoLE_inter_CKM_respond}

\begin{thebibliography}{10}
\providecommand{\url}[1]{#1}
\csname url@samestyle\endcsname
\providecommand{\newblock}{\relax}
\providecommand{\bibinfo}[2]{#2}
\providecommand{\BIBentrySTDinterwordspacing}{\spaceskip=0pt\relax}
\providecommand{\BIBentryALTinterwordstretchfactor}{4}
\providecommand{\BIBentryALTinterwordspacing}{\spaceskip=\fontdimen2\font plus
\BIBentryALTinterwordstretchfactor\fontdimen3\font minus
  \fontdimen4\font\relax}
\providecommand{\BIBforeignlanguage}[2]{{%
\expandafter\ifx\csname l@#1\endcsname\relax
\typeout{** WARNING: IEEEtran.bst: No hyphenation pattern has been}%
\typeout{** loaded for the language `#1'. Using the pattern for}%
\typeout{** the default language instead.}%
\else
\language=\csname l@#1\endcsname
\fi
#2}}
\providecommand{\BIBdecl}{\relax}
\BIBdecl

\bibitem{ZengYong_2021_IEEEWC}
Y.~Zeng and X.~Xu, ``Toward environment-aware 6{G} communications via channel
  knowledge map,'' \emph{IEEE Wireless Commun.}, vol.~28, no.~3, pp. 84--91,
  2021.

\bibitem{ZengYong_2024_IEEE_ST}
Y.~Zeng, J.~Chen, J.~Xu, D.~Wu, X.~Xu, S.~Jin, X.~Gao, D.~Gesbert, S.~Cui, and
  R.~Zhang, ``A tutorial on environment-aware communications via channel
  knowledge map for 6$\text{G}$,'' \emph{IEEE Commun. Surveys Tuts.}, pp. 1--1,
  2024, doi: {\color{blue} \href{http://dx.doi.org/10.1109/COMST.2024.3364508}
  {10.1109/COMST.2024.3364508}}.

\bibitem{JieXu_2022_WCNC}
H.~Li, P.~Li, G.~Cheng, J.~Xu, J.~Chen, and Y.~Zeng, ``Channel knowledge map
  ({CKM})-assisted multi-uav wireless network: {CKM} construction and {UAV}
  placement,'' \emph{J. Commun. Inf. Netw.}, vol.~8, no.~3, pp. 256--270, 2023.

\bibitem{ChenJunting_2023_TWC}
W.~Liu and J.~Chen, ``{UAV}-aided radio map construction exploiting environment
  semantics,'' \emph{IEEE Trans. Wireless Commun.}, vol.~22, no.~9, pp.
  6341--6355, 2023.

\bibitem{Rizk_1997_TVT}
K.~Rizk, J.-F. Wagen, and F.~Gardiol, ``Two-dimensional ray-tracing modeling
  for propagation prediction in microcellular environments,'' \emph{IEEE Trans.
  Veh. Technol.}, vol.~46, no.~2, pp. 508--518, 1997.

\bibitem{Malmirch_2012_TWC}
M.~Malmirchegini and Y.~Mostofi, ``On the spatial predictability of
  communication channels,'' \emph{IEEE Trans. Wireless Commun.}, vol.~11,
  no.~3, pp. 964--978, 2012.

\bibitem{ZengYong_2022_WCNC}
K.~Li, P.~Li, Y.~Zeng, and J.~Xu, ``Channel knowledge map for environment-aware
  communications: E$\text{M}$ algorithm for map construction,'' in \emph{2022
  IEEE Wireless Commun. Netw. Conf. (WCNC)}, 2022, pp. 1659--1664.

\bibitem{XiaoliXu_2024_TWC}
X.~Xu and Y.~Zeng, ``How much data is needed for channel knowledge map
  construction?'' \emph{IEEE Trans. Wireless Commun.}, pp. 1--1, 2024, doi:
  {\color{blue} \href{http://dx.doi.org/10.1109/TWC.2024.3397964}
  {10.1109/TWC.2024.3397964}}.

\bibitem{Krige_2017_TVT}
H.~Braham, S.~B. Jemaa, G.~Fort, E.~Moulines, and B.~Sayrac, ``Fixed rank
  kriging for cellular coverage analysis,'' \emph{IEEE Trans. Veh. Technol.},
  vol.~66, no.~5, pp. 4212--4222, 2017.

\bibitem{KNN_2009_TIP}
K.~S. Ni and T.~Q. Nguyen, ``An adaptable $k$ -nearest neighbors algorithm for
  mmse image interpolation,'' \emph{IEEE Trans. Image Process.}, vol.~18,
  no.~9, pp. 1976--1987, 2009.

\bibitem{IDW_2012_ISCT}
D.~Denkovski, V.~Atanasovski, L.~Gavrilovska, J.~Riihijärvi, and P.~Mähönen,
  ``Reliability of a radio environment map: Case of spatial interpolation
  techniques,'' in \emph{2012 7th Int. ICST Conf. on Cognitive Radio Oriented
  Wireless Netw. and Commun. (CROWNCOM)}, 2012, pp. 248--253.

\bibitem{Naranjo_2014_WCNC}
J.~D. Naranjo, A.~Ravanshid, I.~Viering, R.~Halfmann, and G.~Bauch,
  ``Interference map estimation using spatial interpolation of {MDT} reports in
  cognitive radio networks,'' in \emph{2014 IEEE Wireless Commun. Netw. Conf.
  (WCNC)}, 2014, pp. 1496--1501.

\bibitem{Levie_2021_TWC}
R.~Levie, c.~Yapar, G.~Kutyniok, and G.~Caire, ``{RadioUNet}: Fast radio map
  estimation with convolutional neural networks,'' \emph{IEEE Trans. Wireless
  Commun.}, vol.~20, no.~6, pp. 4001--4015, 2021.

\bibitem{CehnGuokai_2023_CL}
G.~Chen, Y.~Liu, T.~Zhang, J.~Zhang, X.~Guo, and J.~Yang, ``A graph neural
  network based radio map construction method for urban environment,''
  \emph{IEEE Commun. Lett.}, vol.~27, no.~5, pp. 1327--1331, 2023.

\bibitem{ChenJunting_2024_ICC}
Z.~Zhang, G.~Zhu, J.~Chen, and S.~Cui, ``Fast and accurate cooperative radio
  map estimation enabled by {GAN},'' in \emph{2024 IEEE Int. Conf. Commun.
  Workshops (ICC Workshops)}, 2024, pp. 1641--1646.

\bibitem{ShibliAli_2024_ICC}
A.~Shibli and T.~Zanouda, ``Data-driven radio environment map estimation using
  graph neural networks,'' in \emph{2024 IEEE Int. Conf. Commun. Workshops (ICC
  Workshops)}, 2024, pp. 650--655.

\bibitem{LiuCong_2021_ICMA}
C.~Liu, X.~Zhou, J.~Wang, Y.~Li, and X.~Shi, ``Multi-modal channel charting by
  integrating semantic knowledge,'' in \emph{2021 IEEE Int. Conf. Mechatronics
  Autom. (ICMA)}, 2021, pp. 1221--1226.

\bibitem{ITU_building_distribution_2012}
``Propagation data and prediction methods required for the design of
  terrestrial broadband radio access systems operating in a frequency range
  from 3 to 60{GHz},'' Int. Telecommun. Unino, Genva, Switzerland, Tech. Rep.,
  ITU-Recommendation P.1410-5, Fed, 2012.

\end{thebibliography}
\end{document}